\documentclass{article}
\usepackage[utf8]{inputenc}
\usepackage{authblk}

\makeatletter
\renewcommand\AB@affilsepx{, \protect\Affilfont}
\makeatother

\usepackage{booktabs}
\usepackage[singlelinecheck=false]{caption}
\DeclareCaptionFormat{custom}
{%
    \textbf{#1#2}{\small #3}
}
\usepackage{colortbl}
\usepackage{float}
\usepackage[top=1in, bottom=0.8in, left=1in, right=1in]{geometry}
\usepackage{graphicx}
\usepackage{hyperref}
\hypersetup{
    colorlinks=true,
    urlcolor=blue,
    linkcolor=black,
    citecolor=blue
    }
\usepackage{multirow}
\usepackage{biblatex}
\usepackage{tabulary}
\usepackage[normalem]{ulem}
\useunder{\uline}{\ul}{}
\newcommand{\absdiv}[1]{%
  \par\addvspace{.5\baselineskip}
  \noindent\textbf{#1}\quad\ignorespaces
}

\title{\huge \textbf {Crowdsourcing Dermatology Images with Google Search Ads: Creating a Real-World Skin Condition Dataset}}
\date{}

\author[1]{\normalsize{Abbi Ward $^*$}}
\author[1]{Jimmy Li $^*$}
\author[1]{Julie Wang}
\author[1]{Sriram Lakshminarasimhan}
\author[1]{Ashley Carrick}
\author[1]{Bilson Campana}
\author[1]{Jay Hartford}
\author[1]{Pradeep Kumar S}
\author[1]{Tiya Tiyasirichokchai}
\author[1]{Renee Wong}
\author[1]{Sunny Virmani}
\author[1]{Yossi Matias}
\author[1]{Greg Corrado}
\author[1]{Dale Webster}
\author[2]{Dawn Siegel}
\author[2]{Steven Lin}
\author[2]{Justin Ko}
\author[1]{Alan Karthikesalingam}
\author[1]{Christopher Semturs}
\author[1]{Pooja Rao}
\affil[1]{\large{Google Research}}
\affil[2]{Stanford University School of Medicine}

\begin{document}
\newgeometry{top=0.6in, bottom=0.6in, left=1in, right=1in} 
\maketitle
\def\thefootnote{}
\begin{NoHyper}
\footnotetext{\hspace{-0.5cm}* Equal contribution}
\footnotetext{\hspace{-0.5cm}Correspondence: pjro@google.com, \{dsiegel4, stevenlin, jmko\}@stanford.edu}
\end{NoHyper}
\def\thefootnote{\arabic{footnote}}
\vspace{-2em}
\begin{abstract}
    \absdiv{Background:} Health datasets from clinical sources do not reflect the breadth and diversity of disease in the real world, impacting research, medical education, and artificial intelligence (AI) tool development. Dermatology is a suitable area to develop and test a new and scalable method to create representative health datasets.
    
    \absdiv{Methods:}We used Google Search advertisements to invite contributions to an open access dataset of images of dermatology conditions, demographic, and symptom information. With informed contributor consent, we describe and release this dataset containing 10,408 images from 5,033 contributions from internet users in the United States over 8 months starting March 2023. The dataset includes dermatologist condition labels as well as estimated Fitzpatrick Skin Type (eFST) and Monk Skin Tone (eMST) labels for the images.
    
    \absdiv{Results:} We received a median of 22 submissions/day (IQR 14–30). Female (66.72\%) and younger (52\% $<$ age 40) contributors had a higher representation in the dataset compared to the US population, and 32.6\% of contributors reported a non-White racial or ethnic identity. Over 97.5\% of contributions were genuine images of skin conditions. Dermatologist confidence in assigning a differential diagnosis increased with the number of available variables, and showed a weak correlation with image sharpness (Spearman’s P values $<$ 0.001 and 0.01 respectively). Most contributions were short-duration (54\% with onset $<$ 7 days ago ) and 89\% were allergic, infectious, or inflammatory conditions. eFST and eMST distributions reflected the geographical origin of the dataset. The dataset is available at \href{https://github.com/google-research-datasets/scin} {github.com/google-research-datasets/scin}. 

    \absdiv{Conclusion:} Search ads are effective at crowdsourcing images of health conditions. The SCIN dataset bridges important gaps in the availability of representative images of common skin conditions.
    
  \end{abstract}
\section*{Introduction}

\subsection*{Need for better ways to create representative, diverse and accessible datasets for health education and research}
Health datasets\cite{irvin2019chexpert,Johnson2016-gf,Rotemberg2021-lz} intended for use in research and medical education are traditionally obtained through prospective clinical studies, or retrospectively from healthcare facility electronic health record (EHR) systems. However, a patient’s healthcare journey begins much earlier than their first encounter with the healthcare system\cite{noauthor_undated-kf}, and datasets derived from clinical sources are unlikely to capture disease presentation

\restoregeometry
 prior to an established diagnosis or signs and symptoms occurring between episodes of care. Datasets of clinical origin also often reflect the underlying disparities in healthcare access across race/ethnicity\cite{Kleinberg2022-ql}, geography\cite{Celi2022-nx}, socioeconomic status\cite{Seyyed-Kalantari2021-kf}, and other dimensions\cite{Wen2022-al}.

\subsection*{Dermatology is uniquely positioned for new dataset creation}

Dermatology is an area that lends itself well to creating a crowdsourced dataset where recruitment is directly centered on the individual experiencing a health condition, rather than on the subset of individuals with access and presenting to the health system. Skin conditions, being on the external surface of the body, are relatively easy for individuals to photograph, and many could be diagnosable by dermatologists\cite{Wallach2005-eq,Sellheyer2005-bl} through a combination of images and self-reported symptoms.  

Dermatology is also an area where the need for representative images for research, medical education, and patient care is evident. Access to dermatologists is lacking in many community, rural, and global settings\cite{Brodell2021-wp,Chang2017-xh}, and primary care physicians manage 70\% of skin disease encounter\cite{Feldman1997-le}. To help with this, many research teams have developed AI-based tools to help interpret images of skin conditions\cite{Han2020-qz,Jain2021-it}. However, most existing dermatology datasets used to develop these AI models over-represent skin malignancies\cite{Wen2022-al,Guo2022-bz} and under-represent darker skin tones \cite{Kleinberg2022-ql,Wen2022-al}.  These biases in the underlying data can lead to fairness or equity gaps in the AI tools\cite{Daneshjou2022-pl}. Further, detecting such issues can be difficult because most dermatology datasets do not report patient race/ethnicity\cite{Daneshjou2021-yz}. 

There have been efforts to remedy the lack of representative dermatology datasets for both clinical users\cite{Daneshjou2022-pl,groh2022transparency} and everyday internet users\cite{noauthor_undated-vh,noauthor_undated-oy}. Yet major unmet needs remain. Few dermatology datasets reflect the breadth and diversity of disease in the real world including common, short-duration, infectious, allergic, or inflammatory conditions. Dataset creation is resource-intensive and the collection of large datasets is out of the reach of most researchers and educators. 

\subsection*{Web search is a key component of modern healthcare journeys}

An alternative to traditional EHR-based data extraction is crowdsourcing directly from patients or users via the internet. Web symptom search is a key component of the modern healthcare journey, with 80\% of US internet users\cite{Fox2011-di} or 55.8\% of US adults\cite{Wang2023-nd} reporting having used the internet to look up health or medical information. Researchers have attempted to study the spread of infectious disease\cite{Ginsberg2009-xk,Walker2020-nx}, detect foodborne illness hubs\cite{Sadilek2018-ea}, or forecast national suicide rates\cite{Barros2019-nm}, from aggregated search queries, and to predict the onset of malignancies\cite{Paparrizos2016-rw,White2017-yu} from individual search history. Internet search queries could be a powerful tool to not only study the evolution of health and disease, but also to reach individuals at early stages of their healthcare concerns.

\subsection*{Investigating the use of Search advertisements to scale recruitment for open-access dataset creation}

Medical researchers have investigated the use of search, social media and other online advertising methods for clinical study recruitment\cite{Upadhyay2020-cf,Akers2018-kz,Anguera2016-sa,Gordon2006-pg}, with promising results. Aside from their reach, the specific feature that renders search ads valuable for dataset creation is that they can be selectively displayed to individuals actively searching for specific symptoms or health-related keywords, representing the population that is seeking that specific health information.

In this work, we use Google Search advertisements (ads) to create a consented dermatology image dataset — the Skin Condition Image Network (SCIN) dataset — composed of voluntary contributions from internet users. After curation and de-identification, we release the dataset as a resource for health education, research, and tool development. 
The study aimed to address the following questions:
\begin{enumerate}
    \item Is crowdsourcing with search ads effective? 
    \item How well does the crowdsourced dataset represent the population?
    \item How do datasets generated with this method differ from those derived from clinical sources?
\end{enumerate}

\section*{Methods}
\subsection*{Search advertisements}
Starting with a list of 100 skin-, hair-, and nail conditions (Table \ref{table:tableS1}), we used the \href{https://ads.google.com/home/tools/keyword-planner/}{Google Search Ads Keyword Planner tool} to generate and select a list of 3000 skin-related search keywords. The keyword list included common keywords like ‘rash pictures’, ‘heat bumps’ or ‘hives on skin’, that were relevant to more than one dermatological condition, as well as more condition-specific terms like ‘mole with irregular border’, ‘plaque psoriasis on feet’, or ‘filiform warts’. We then displayed text ads inviting contributions to the SCIN open access dataset selectively to adults in the United States (US) searching for these keywords on a mobile device (to facilitate photo taking). There was no demographic or additional geographic selection or exclusion. Ads had under 10\% impression share for each keyword, meaning that fewer than 10\% of individuals who searched for a keyword would see a SCIN ad. 

The keyword selection, text ads display,  and measurement tools used in this study are standard Google Ads features, available for general use. Similar to prior health-related studies that use Google Search Ads for recruitment\cite{Upadhyay2020-cf,Gordon2006-pg} we used non-personalized advertising — meaning that the ads did not use or imply any prior knowledge about users’ health.

\subsection*{SCIN web application}
Individuals who clicked on the advertisement reached the landing page of a web application (i.e., within their browser, without installing a separate app, Figure \ref{fig:figS1}) explaining the crowdsourcing initiative, the motivation for the open access dataset, and who was eligible to participate. Participation was limited to individuals above 18 and in the US. Those who clicked ‘Get Started’, went through an online IRB-approved informed consent process, and contributed 1-3 images of their skin condition. The application requested 1 close-up photograph, 1 photograph from an angle and 1 photograph from a distance showing the entire affected body part. Images could be either uploaded from an existing photo library or taken live. Contributors were then given the option to answer a series of questions about their demographic information (age, sex at birth, and race/ethnicity), and provide information related to their skin condition, including self-reported Fitzpatrick Skin Type (FST) based on tanning/burning propensity. Information was collected through a series of checkboxes and drop-down menus as shown in Table \ref{table:tableS2}.

This study was approved by the Advarra Institutional Review Board (Pro00068073).

\subsection*{Data processing}
After upload, a third-party team of trained personnel de-identified each image by redacting any areas that could identify individuals, such as facial areas, tattoos, or landmarks in the image. Each de-identified image went through a further quality check to make sure that any identifying information was removed. 

We then filtered the dataset by deleting the following categories:
\begin{enumerate}
\item Spam or images not showing a skin condition (manual, marked during the de-identification process described above)
\item Duplicates (algorithmic, using computed image features)
\item Image Safety: Images that could show children (algorithmic, using a ‘child presence’ score\cite{Jasper2022-or})
\item Image Safety: Images that could contain pornographic content (algorithmic, using a combination of ‘pornographic’ and ‘medical image’ scores\cite{noauthor_undated-mm} to filter out potentially unsafe images while retaining medically relevant ones)
\item Images where no skin condition was visible after redaction (manual)
\end{enumerate}

Subsequently, each submission (images with demographic and symptom information) was evaluated by 1–3 dermatologists out of a pool of 10 US board-certified dermatologists (mean years of experience: 14.8, range: 5–35). The first 750 submissions with self-reported health and demographic information were reviewed by 3 dermatologists and the remaining were reviewed by 1 dermatologist. Dermatologists provided a differential diagnosis and a confidence score from 1–5 for each condition label in the differential diagnosis, using a list of 419 conditions created using SNOMED CT condition categories and a labeling interface, as previously described\cite{Liu2020-oc}. Dermatologists also estimated Fitzpatrick Skin Type based on the images (eFST). The differential diagnoses for each submission were mapped into a list of condition labels and weights, using a previously described aggregation procedure\cite{Liu2020-oc}. Submissions that were not diagnosable are included in the SCIN dataset, and labeled as such.  Each submission was also labeled with estimated Monk Skin Tone\cite{Monk2023-cf} (eMST) by 2 trained, non-clinical annotators, from separate geographical locations(US and India). The weighted differential diagnosis of the skin condition, median eFST and eMST labels are released with the dataset.

\subsection*{Measures to protect contributor privacy}
Volunteers were fully informed that their image contributions would be included in an open dataset; this information was provided through ad copy, landing page text, FAQ sections, and the informed consent document. Through the consent and image upload process, contributors were warned about the risk of potential re-identification, and advised not to upload images containing identifying features. They were also advised not to upload the same images on social media or other sites.

In addition to the manual image de-identification procedure described under ’Contributor data processing’, we passed the images through a reverse web search to exclude any contributions with near-identical copies publicly available. We scrubbed image (EXIF) metadata, and redacted areas on the image containing text through a partially automated optical character reading (OCR) and masking process. 

Contributor age was bucketed in 10-year intervals. Where there were fewer than 10 individuals for any combination of potentially identifying variables, we masked these variables by converting them to ‘unknown’, or bucketed them into a larger group. 

The SCIN Data Use License prohibits users of the open access data from attempting to re-identify individuals that contributed to the SCIN dataset.

\subsection*{Data analysis and reporting}

\subsubsection*{Ads}
We obtained ad impressions, click-through rates, and submission rates, and their age and gender breakdown, from the advertiser-facing Google Search ads dashboard for the duration of the campaign. We report these as monthly averages.

\subsubsection*{Dataset}
Image quality scores were computed using standard photograph sharpness and exposure quality metrics designed for regular images\cite{noauthor_undated-gc}. The scores are visualized across all images in the dataset. For analysis, we use the image sharpness and exposure quality thresholds in (Table \ref{table:tableS3}). As the scores were computed after redacting identifying features, the effects of the redaction can be seen on the score. For example, the presence of straight lines demarcating redacted areas causes these images to have high sharpness scores. When analyzing dermatologist confidence against image quality, we used the average score across all images in the submission, and the top dermatologist confidence across their differential. 

The analysis of age, sex, and race/ethnicity versus US population averages excluded cases where the contributor chose ‘Prefer not to answer’. When choosing public datasets for comparison of disease distribution, we included those where information about diagnosis or diagnostic category was available with numbers for each diagnosis or category. When choosing public datasets for comparison of FST and eFST, we included datasets that had not been enriched for population groups, geographies or skin types. Where FST was reported even for a fraction of the dataset\cite{Wen2022-al}, we included those numbers as representative of the entire dataset.

\section*{Results}
\subsection*{Search ads are an effective method for crowdsourcing consented images of dermatology conditions}

\begin{figure}[ht!]
\caption{A: Funnel from ad view to submission to the SCIN dataset. B, C: Monthly ad impressions, click-through rates, and submission rates by gender and age}
\label{fig:fig1}
\includegraphics[width=\textwidth]{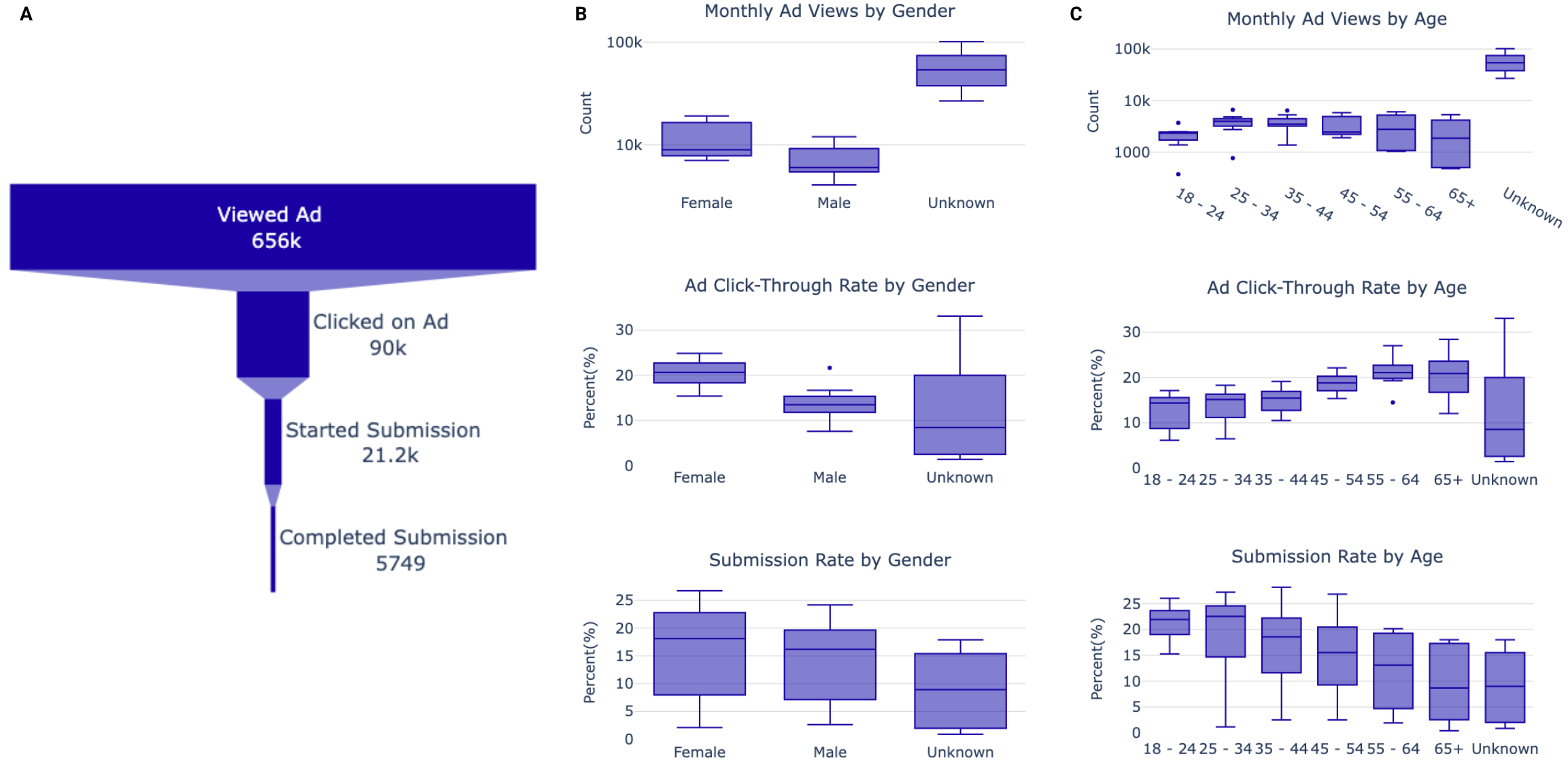}
\end{figure}

Starting March 2023, ads for contributing to the SCIN dataset were displayed to individuals using Google Search in the United States searching for keywords related to dermatology conditions. Of the individuals to whom an ad was displayed, 13.7\% (90,445/65,6674) clicked on it; 6.4\% (5,749/90,445) of those who clicked on the ad consented to providing images and completed the submission process (Figure \ref{fig:fig1}A). Overall, the SCIN dataset received a median of 22 submissions/day (IQR 14-30), with a total of 5,749 submissions over 251 days. The volume of contributions did not vary substantially over time.

Based on data from the Google Search ads tool, significantly more female than male individuals viewed a SCIN ad (Figure \ref{fig:fig1}B, Table \ref{table:tableS4}). Female individuals were also significantly more likely to click through to the application landing page (Figure \ref{fig:fig1}B, Table \ref{table:tableS4}). Having clicked on the ad, female and male individuals were similarly likely to make a submission to the dataset. Monthly ad views were similar across age groups. Older individuals were significantly more likely to click on an ad (Figure \ref{fig:fig1}C, Table \ref{table:tableS4}). Having reached the landing page of the application, younger individuals were more likely to complete the process of contributing to the dataset (Figure \ref{fig:fig1}C, Table \ref{table:tableS4}).

\subsection*{Individuals submit high-quality photographs that dermatologists are able to label with a differential diagnosis}
\begin{figure}[H]
\caption{ A: Sankey diagram illustrating dropout at each step of the data curation process for the SCIN dataset B: Distribution of image sharpness and exposure quality scores C: Dermatologist label confidence by image exposure quality and sharpness score D: Dermatologist label confidence by number of self-reported variables E: A sample of images from the SCIN dataset with dermatologist labels and label confidence}
\label{fig:fig2}
\includegraphics[width=\textwidth]{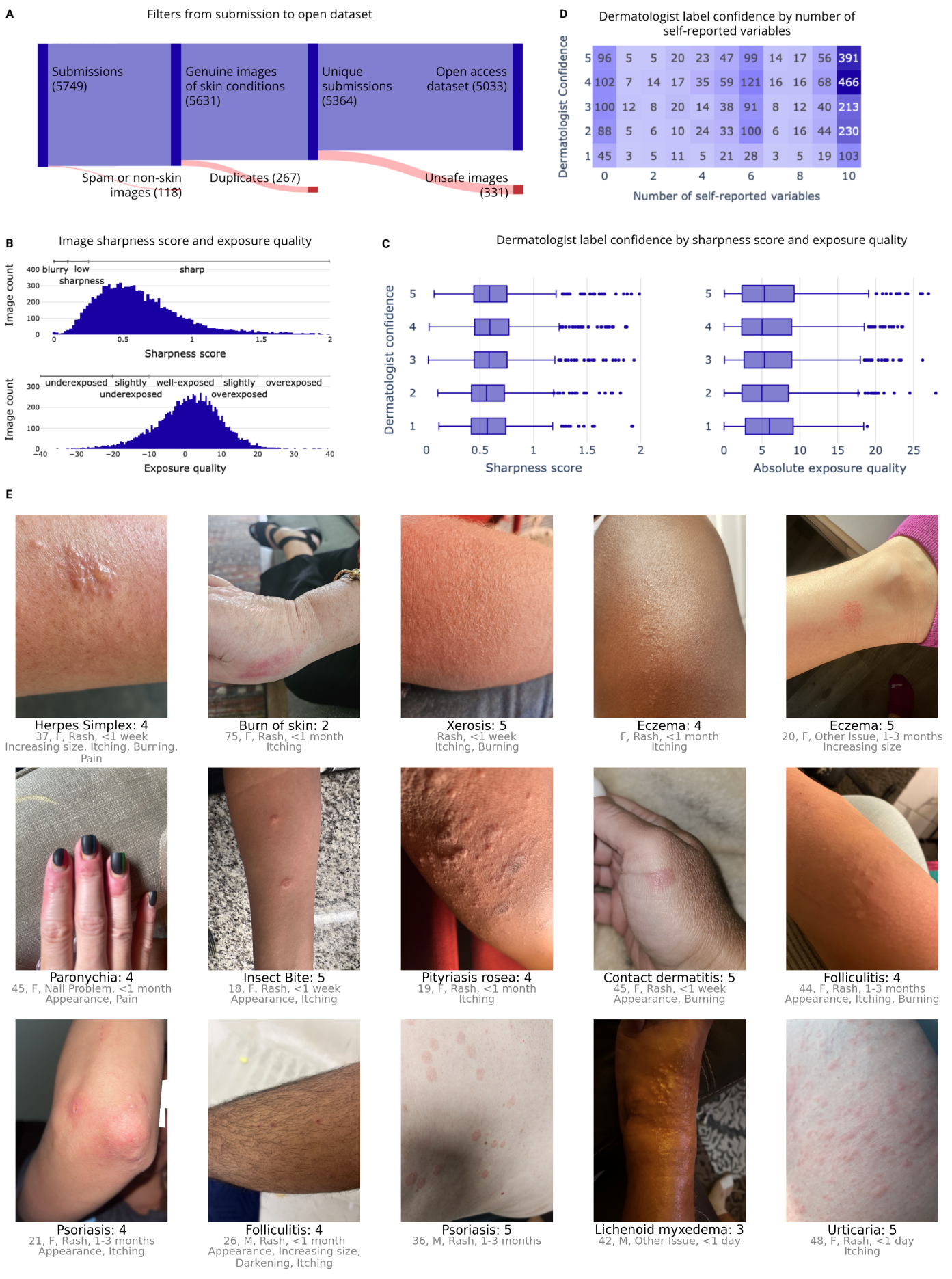}
\end{figure}

Over 97.5\% of submissions (5,631/5,749) contained genuine images of skin conditions; 118 of 5,749 submissions were spam or did not contain images of a skin condition (Figure \ref{fig:fig2}A). The only further filtering performed was to exclude duplicates across submissions (N=267) and for image safety reasons (N=331). 

Most images were well-exposed and sharp (Figure \ref{fig:fig2}B). Dermatologist confidence was weakly correlated with image sharpness (Figure \ref{fig:fig2}C; Spearman’s R [p value]: 0.054 [0.01]), and not correlated with absolute image exposure quality (Spearman’s R [p value]: 0.007 [0.96]). Dermatologists' ability to provide a differential diagnosis — or image diagnosability — varied from 93\% (614/658) for images with 10 self-reported variables viewed by 3 dermatologists, to 59\% (452/767) for images that were viewed by a single dermatologist and had no self-reported demographic or symptom information. Similarly, dermatologists’ confidence in their differential diagnosis was correlated with the number of self-reported variables provided (Figure \ref{fig:fig2}D, Spearman’s R [p value]: 0.15 [$<$0.001]).

\subsection*{Women and younger individuals are more highly represented}
\begin{figure}[ht!]
\caption{A: Dropout from image submission to inclusion in the open access SCIN dataset. B-D: SCIN dataset breakdown by age, sex and race/ethnicity* compared with United States Census population estimates}
\label{fig:fig3}
\includegraphics[width=\textwidth]{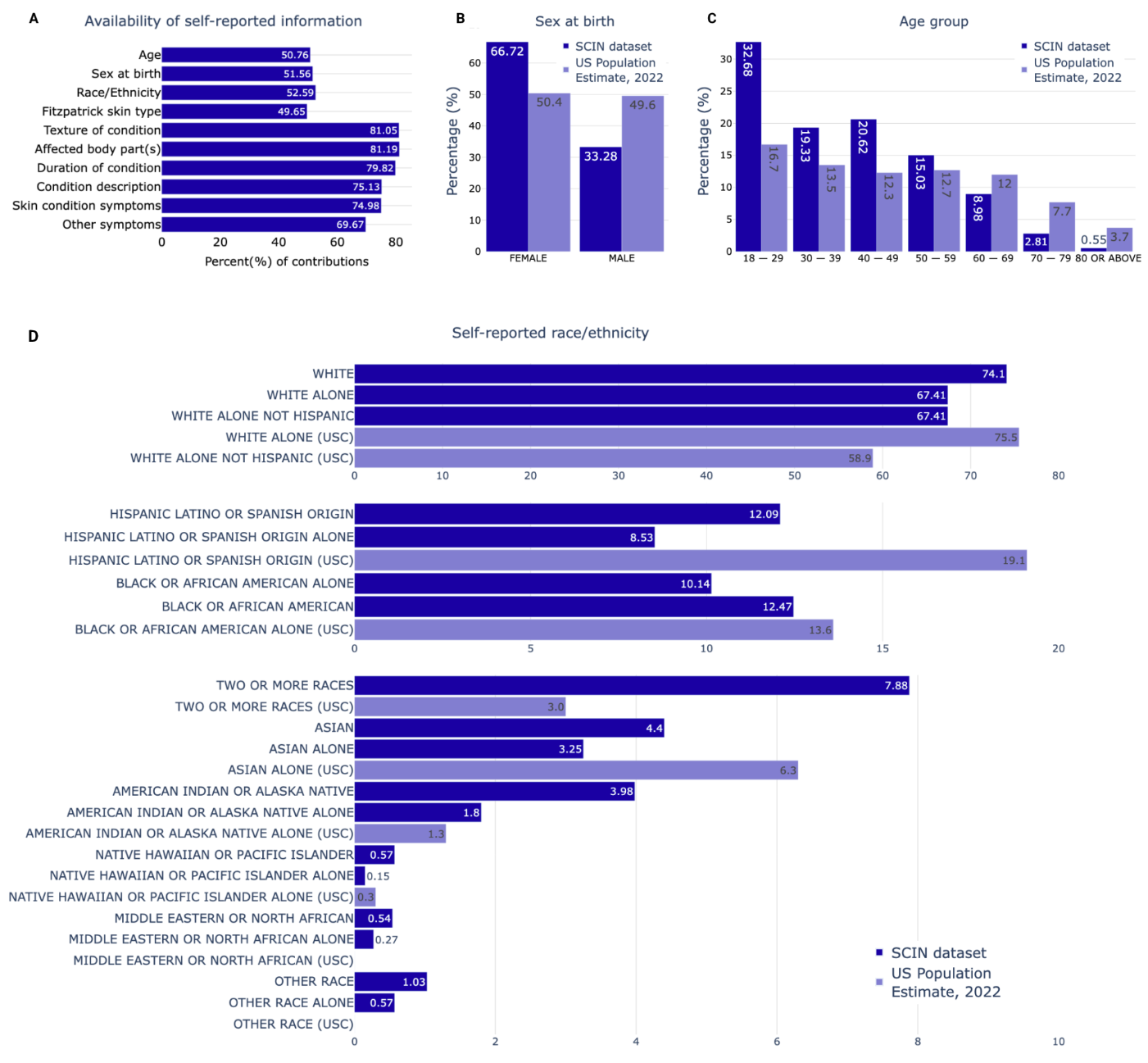}
\end{figure}

Contributors submitted 1–3 images (average 2.06) of their skin concern. Self-report of demographic and symptom information was optional; approximately 50\% of contributors reported age, sex at birth, and race/ethnicity, and approximately 70\% or more reported symptom information (Figure \ref{fig:fig3}A, Table \ref{table:tableS5}). Contributors were most likely to report the affected body part and texture of the condition, and least likely to assess and report their Fitzpatrick skin type (self-reported FST or sFST). 

Based on data from the 50\% of contributors who reported demographic information, female individuals had significantly higher representation (66.72\%, 1724/25844, chi-square p value $<$0.001, Figure \ref{fig:fig3}B, Table \ref{table:tableS6}) in the dataset compared to the US population census estimates\cite{United_States_Census_Bureau_Communications_Directorate-_Center_for_New_Media_undated-dz}. Younger individuals also had higher representation relative to the census (Figure \ref{fig:fig3}C, Table \ref{table:tableS6}). 32.6\% (852/2614) of contributors reported having a non-White racial or ethnic identity. The self-reported racial and ethnic distribution is difficult to compare directly with US population estimates\cite{United_States_Census_Bureau_Communications_Directorate-_Center_for_New_Media_undated-dz}, due to differences in the way that the question was asked (see Methods) and the number of categories. However, individuals identifying as American Indian or Alaska Native, Middle Eastern or North African, ‘Other Race’, or ‘Two or more races’ had clearly higher representation compared to the population average (Figure \ref{fig:fig3}D, Table \ref{table:tableS6}), and Asian, Hispanic, and African-American groups had lower representation.

\subsection*{The SCIN dataset contains a range of common, recent-onset skin conditions, across Fitzpatrick skin types}

\begin{figure}[H]
\caption{A: Percentage distribution of dermatological condition categories in the SCIN dataset compared with existing publicly available datasets B: Self-reported (sFST) and dermatologist-labeled (eFST) Fitzpatrick skin type percentage distribution of the SCIN dataset compared with publicly available datasets C: Estimated Monk Skin Tone (eMST) percentage distribution of the SCIN dataset, showing variation in eMST labels for the same set of images between labelers from two geographies}
\label{fig:fig4}
\includegraphics[width=\textwidth]{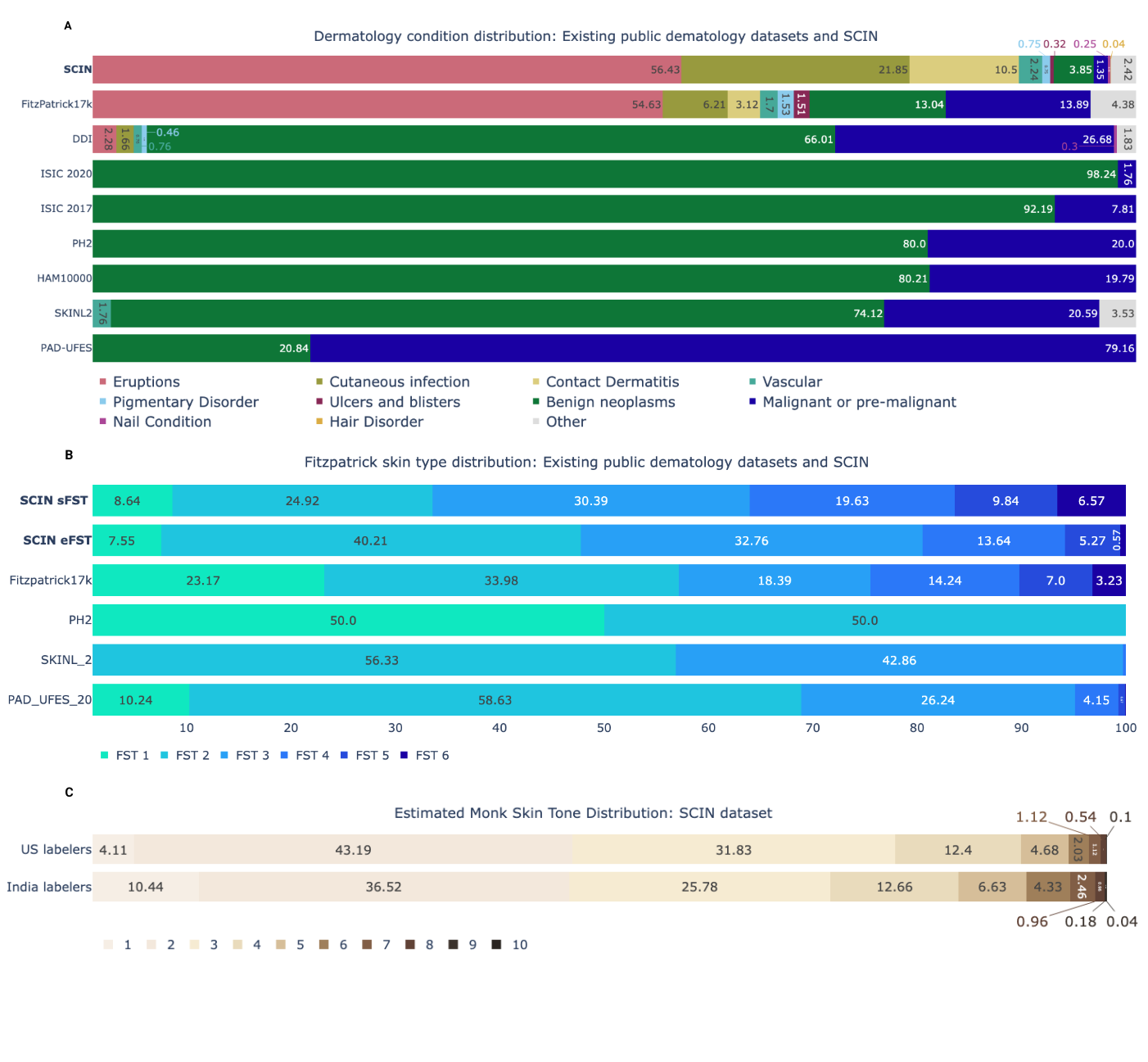}

\end{figure}

Based on retrospective dermatologist labeling of the crowdsourced dataset, the submitted images are of largely common, recent-onset conditions of inflammatory and infectious origin (Figure \ref{fig:fig4}A). Nearly 2\% of contributions represented concerns arising within the past day, 54\% within the past week, and 78\% within the past month (Table \ref{table:tableS7}). Neoplasms comprise less than 5\% of the SCIN dataset, while eruptions (rashes), cutaneous infections, and contact dermatitis represent over 70\% (Figure \ref{fig:fig4}A, Table \ref{table:tableS8}). 

This composition complements existing datasets\cite{Rotemberg2021-lz,Wen2022-al,Tschandl2018-wc,Pacheco2020-kr,De_Faria2019-wi,Mendonca2013-ny,Groh2021-wg,Stoyanov2018-cs} from clinical sources, which largely focus on benign and malignant neoplasms Figure \ref{fig:fig4}A, Table \ref{table:tableS9}). 

Crowdsourcing images of skin conditions from US internet search users, without selectively enriching population groups, results in a dataset that is broadly distributed across Fitzpatrick skin type (eFST and sFST) compared to existing datasets that are similarly un-enriched for skin types, geography or population subgroups (Figure \ref{fig:fig4}B, Table \ref{table:tableS10}). sFST differed from eFST for many images (Figure \ref{fig:figS2}). The eMST labels in the dataset (Figure \ref{fig:fig4}C, Table \ref{table:tableS11}) provide a more nuanced representation of darker skin tones than eFST, showing that a US-origin dataset un-enriched for skin tone or type has less than 1.5\% of individuals with eMST 8, 9, or 10. eMST rating varied with rater geography, resulting in two different eMST distributions for the same set of images (Figure \ref{fig:fig4}C).

\section*{Discussion}
\subsection*{Benefits and limitations of crowdsourcing with search ads}
Through this prospective study, we demonstrate the feasibility of using search advertisements to crowdsource images of dermatology conditions, and to create a representative, diverse, and high-quality consented health image dataset. In contrast to most existing dermatology datasets that are designed for malignancy detection\cite{Daneshjou2022-pl}, the SCIN dataset contains largely common, short-duration, allergic,  inflammatory, and infectious conditions. This dataset distribution, along with the range of body parts, skin types, and demographics represented, is a key strength of the crowdsourcing method. 

Using web search as the entry point for crowdsourcing allows researchers to reach individuals at earlier stages of their skin concern — often before they have sought clinical care — and to include individuals who may face barriers in accessing healthcare services. The spectrum of skin concerns in this dataset thus provides a valuable supplement to that seen in medical textbooks, atlases, or datasets from EHRs. 

We believe that the conditions represented in the SCIN dataset approximately represent the top dermatology concerns of internet users in the US. A more rigorous study of the prevalence of skin conditions in the population of internet users, including seasonal trends, would require controlled experiments with search keywords and submission volumes (Future work, Table \ref{table:table1}). 

Limitations of the crowdsourcing method are similarly reflected in the dataset. The dataset contains few images of rarer conditions, and chronic conditions are under-represented because individuals who have already received a diagnosis tend to search for prognosis and care recommendations\cite{Orgaz-Molina2015-dr}. These constraints could be overcome through keyword expansion if consistent with the goals for dataset collection (e.g., to help interpret images of a new concern vs monitoring progression of an existing condition).

\subsection*{Characteristics of the SCIN dataset}
The higher representation of younger individuals and female individuals in the dataset is explained through a combination of ‘who has dermatological concerns’,  ‘who searches for dermatology conditions using mobile devices’, ‘who clicks on ads’, and ‘who is likely to volunteer to upload images for an open access dataset’. On average, the incidence of skin and subcutaneous conditions is higher for female individuals and younger patients\cite{Yakupu2023-nm}. The higher rate of online health information seeking among women compared to men has been also observed before across geographies and cultural settings\cite{Wang2023-nd,Renahy2008-lq,Ghweeba2017-zz,Bundorf2006-vr}. While the self-reported racial/ethnic categorization we used was not directly comparable with US census methods, the rate of minority group participation in the creation of the dataset is encouraging. 

Race or ethnicity is seldom reported in dermatology datasets\cite{Wen2022-al}. We hope that including this self-reported information facilitates more research on fairness and equity in dermatology, as well as into the types and manifestations of clinical conditions among various population groups. 

We provide retrospectively obtained dermatologist skin condition labels for the dataset. Unlike malignancies which require biopsy confirmation\cite{Nelson2019-ee}, many common conditions of the kind seen in the SCIN dataset lend themselves well to image-based clinical labeling\cite{Nelson2016-bg}. Although dermatologists had high confidence in being able to provide a differential diagnosis for many submissions, these labels should not be considered equivalent to a clinical diagnosis. The lack of correlation between image quality and dermatologist labeling confidence could be due to our use of standard image quality metrics rather than ‘medically appropriate’ image quality\cite{Vodrahalli2023-qj}, or because the labeling dermatologists had access to more context in the form of self-reported information.

The choice between skin type (functional status incorporating tanning propensity and cancer risk) and skin tone (based on shade matching) depends on use case, and these labels in the SCIN dataset should be interpreted with care. Fitzpatrick skin type evaluation, whether through self-assessment\cite{Rampen1988-mp} or retrospective eFST labeling from photographs\cite{Groh2022-gh} has known limitations. Similarly, the MST scale was originally devised for use as color comparison cards held directly against the skin\cite{Monk2023-cf}, and eMST can vary based on the labeler geographic origin (Figures \ref{fig:fig4}C, \ref{fig:figS2}). Retrospective labeling of skin type and tone is inherently subjective due to several factors, including variations in inferred lighting, and annotator lived experiences and judgment of skin/eye/hair color. Despite these limitations, the SCIN dataset allows a comparison of sFST, eFST, and eMST for dermatology images (Figure \ref{fig:figS2}),  and these labels are intended to facilitate further dermatology research into skin tone and skin type (Future work, Table \ref{table:table1}). 

In the absence of existing FST or MST distribution estimates in the US for comparison, the SCIN dataset sets an approximate benchmark for the US population distribution of these skin type and tone scales. Un-enriched for demographic group or skin tone, the crowdsourced SCIN dataset is broadly distributed across FST, but has relatively few contributors with MST 8+. This indicates that representing the full MST spectrum would require larger, more global dataset creation or an enriched sampling approach.

\subsection*{Future Work}
The use of search ads for crowdsourcing could potentially be applied to other kinds of health data, especially those that internet users can be expected to have on hand, and for data that can be either self-labeled\cite{Bhattacharya2023-bl}, or labeled retrospectively (Table \ref{table:table1}). By adapting the search keywords and advertising copy, the method could conceivably also be used for rare disease registries, to enhance the reach and diversity of clinical trial recruitment, or for fully digital clinical studies.

Most published AI algorithms for dermatology are trained and evaluated using data from clinical sources, and focus on differentiating malignant from benign lesions\cite{Daneshjou2021-yz}. Even in cases where models are designed to operate across other skin conditions\cite{Liu2020-oc,Son2021-kx}, they are usually not tested using images obtained directly from patients. The SCIN dataset could help evaluate the performance of these tools in the real world (Future work, Table \ref{table:table1}).

\renewcommand{\thetable}{\arabic{table}}
\begin{table}[H]
\caption{Examples of potential future work with the crowdsourcing method and the SCIN dataset}
\label{table:table1}
\begin{tabular}{@{}p{5cm}|p{11cm}@{}}
\toprule
\multirow{3}{4cm}{Health education} & Representative and diverse images for dermatology textbooks, atlases, and online material          \\ \cmidrule(l){2-2}
& Database for training medical students, residents, and practicing providers           \\ \cmidrule(l){2-2}
& Patient-facing educational material        \\ \bottomrule
\multirow{3}{4cm}{Health and population health research} & Modeling or monitoring seasonal trends and outbreaks of dermatological or other health conditions           \\ \cmidrule(l){2-2}
& Prevalence of dermatology conditions in the community versus in health systems           \\ \cmidrule(l){2-2}
& Skin type and skin tone representation in dermatology     \\ \bottomrule
\multirow{4}{4cm}{Consented, population-based health dataset creation} & Rare disease registry           \\ \cmidrule(l){2-2}
& Continuous glucose, sleep, step counts, or heart rate using fitness applications or phone sensors           \\ \cmidrule(l){2-2}
& Cough, speech or vocal markers of health and disease           \\ \cmidrule(l){2-2}
& Environmental signals for health such as air or water quality     \\ \bottomrule
\multirow{4}{4cm}{Health AI research and development} & Evaluating the accuracy and fairness of AI models           \\ \cmidrule(l){2-2}
& Evaluating generalizability across the distribution shift from clinical population to internet users          \\ \cmidrule(l){2-2}
& Self-supervised model training, generative model training           \\ \cmidrule(l){2-2}
& Training and evaluation of models for skin tone or skin type classification     \\ \bottomrule
\end{tabular}
\end{table}

These potential applications in Table \ref{table:table1} point to the method's potential to accelerate healthcare research across various domains. Scaling the crowdsourcing approach responsibly will require attention to consent procedures, robust privacy preservation strategies tailored to specific data types, and clear and accessible communication to engage participants as active contributors to research. At the same time, researchers should recognize that crowdsourced information complements, rather than replaces, clinically sourced data and may lack vital clinical context. Not all data types are suited to retrospective labeling, and alternative strategies may be required if clinical-grade labels are essential to the use case. While the SCIN dataset demonstrates how crowdsourcing can yield greater diversity than some clinical datasets, strategies such as targeted outreach or collection campaigns for underrepresented groups and rarer conditions may be needed to combat potential biases in other domains or at larger scales.

\section*{Conclusion}
Crowdsourcing is an effective and accessible alternative to existing health-system-centric dataset creation methods. We demonstrate that it is possible to create a diverse and representative health dataset without demographic targeting or selective sampling. The SCIN dataset and the use of web search as an entry point open up new possibilities for research, medical education, and AI tool development.

\subsection*{Acknowledgements}
We thank Yetunde Ibitoye, Anna Iurchenko, Sami Lachgar, Lisa Lehmann, Javier Perez, Margaret-Ann Smith, Rachelle Sico, Amit Talreja, Annisah Um’rani and Wayne Westerlind for their essential contributions to this work. We are grateful to Heather Cole-Lewis, Naama Hammel, Ivor Horn, Michael Howell, Yun Liu, and Eric Teasley for their insightful comments on the study design and manuscript.

{\footnotesize
\printbibliography

\section*{Supplementary Material}
\subsection*{Supplementary Tables}
\setcounter{table}{0}
\renewcommand{\thetable}{S\arabic{table}}
\begin{table}[H]
\caption{List of dermatology conditions for keyword generation}
\label{table:tableS1}
\begin{tabular}{|p{16cm}|}
\toprule
  Abscess, Acanthosis nigricans, Acne vulgaris, Acne keloidalis, Actinic keratosis, Allergic contact dermatitis, Alopecia areata, Amyloidosis of skin, Androgenetic alopecia, Angioma, Angular cheilitis, Atypical nevus, Bed bug bites, Bacterial skin rash, Basal cell carcinoma, Blue nevus, Burn, erysipelas, Carbuncle, Cellulitis, Cherry angioma, Chigger bites, Comedone, Cold sore, Cutaneous sarcoidosis, Cyst, Cystic acne, Decubitus ulcer, Dermatitis herpetiformis, Dermatofibroma, Diabetic ulcer, Drug rash, Dyshidrotic eczema, Eczema, Erythema multiforme, Erythema nodosum, Erythrasma, Foliculitis, Flea bites, Freckles, Fungal acne, Genital herpes, Granuloma annulare, Herpes simplex, Herpes zoster, Hidradenitis suppurativa, Hives, Hyperhidrosis, Hyperpigmentation, Ichthyosis vulgaris, Impetigo, Inflicted skin lesions, Ingrown hair, Intertrigo, Irritant contact dermatitis, Keloid, Keratosis pilaris, Lentigo, Lichen planus, Lichen sclerosus, Lipoma, Melanocytic Nevus, Melanoma, Melasma, Milia, Miliaria, Molluscum contagiosum, Nummular dermatitis, Onychomycosis, Papilloma, Paronychia, Perioral dermatitis, Petechiae, Photodermatitis, Pityriasis alba, Pityriasis rosea, Poison ivy rash, Prurigo Nodularis, Psoriasis, Pustular psoriasis, Purpura, Rosacea, Scabies, Scar conditions, Seborrheic Keratosis, Spider bite, Squamous Cell Carcinoma, Stasis dermatitis, Stevens-Johnson syndrome, Sunburn, Tick Bites, Tinea, Tinea Barbae, Tinea Cruris, Tinea Pedis, Tinea versicolor, Urticaria, Verruca Vulgaris, Vitiligo, Viral exanthem\\
\bottomrule
\end{tabular}
\end{table}

\begin{table}[H]
\caption{Contributor questions and possible answers}
\label{table:tableS2}
\centering
\begin{tabular}{p{6cm}|p{9.5cm}}
\toprule
\textbf{Question} & \textbf{Answer}                     \\ \midrule
\textbf{Sex at Birth}                    &Male \textbar\  Female \textbar\ Other \textbar\ Prefer not to say           \\ \midrule
\textbf{Age}                    &Select number from 18 to 120               \\ \midrule
\textbf{With which racial or ethnic groups do you identify? (Mark all boxes that apply)}         & American Indian or Alaska Native \textbar\ Asian \textbar\ Black or African American \textbar\ Hispanic, Latino, or Spanish Origin \textbar\ Middle Eastern or North African \textbar Native Hawaiian or Pacific Islander \textbar\ White \textbar\ Another race or ethnicity not listed \textbar\ Prefer not to answer                 \\ \midrule
\textbf{How does your skin react to sun exposure? (Select one)}  &Always burns, never tans \textbar\ Usually burns, lightly tans \textbar\ Sometimes burns, tans evenly \textbar\ Rarely burns, tans well \textbar\ Very rarely burns, easily tans \textbar\ Never burns, always tans                       \\ \midrule
\textbf{Describe how the affected area feels* (Select all that apply)}  & Raised or bumpy \textbar\ Flat \textbar\ Rough or flaky \textbar\ Filled with fluid                      \\ \midrule
\textbf{How long has this skin issue been going on? (Select one)}  &One day \textbar\ Less than one week \textbar\ One to four weeks \textbar\ One to three months \textbar\ Three months \textbar\ Three to twelve months \textbar\ More than one year \textbar\ More than five years \textbar\ Since childhood \textbar\ Since birth                       \\ \midrule
\textbf{Where on your body is the issue? (Select all that apply)
}   &Head / Neck \textbar\ Arm \textbar\ Palm of hand Back of hand \textbar\ Front torso \textbar\ Back torso \textbar\ Genitalia / Groin \textbar\ Buttocks \textbar\ Leg \textbar\ Top / side of foot \textbar\ Sole of foot \textbar\ Other                       \\ \midrule
\textbf{How would you best describe your current skin concern? (Select one)}  & Acne \textbar\ Growth or mole \textbar\ Hair loss \textbar\ Other hair problem \textbar\ Nail problem \textbar\ Pigmentary problem \textbar\ Rash \textbar\ Other                      \\ \midrule
\textbf{Are you experiencing any of the following with your skin concern? (Select all that apply)}  & Bothersome in appearance \textbar\ Bleeding \textbar\ Increasing in size \textbar\ Darkening \textbar\ Itching \textbar\ Burning \textbar\ Painful                       \\ \midrule
\textbf{Do you have any of these symptoms?}  &Fever \textbar\ Chills \textbar\ Fatigue \textbar\ Joint pain \textbar\ Mouth sores \textbar\ Shortness of breath \textbar\ No relevant symptoms \\ \bottomrule
\end{tabular}
\vspace{0.1cm}
\begin{flushleft} 
\small *Diagrams were shown to illustrate what these words mean
\end{flushleft} 
\end{table}

\begin{table}[H]
\caption{Image sharpness and exposure quality thresholds}
\label{table:tableS3}
\begin{tabular}{@{}l|l|l@{}}
\toprule
\multirow{3}{*}{\textbf{Sharpness score}}  & \textless 0.1     & Blurry                \\ \cmidrule(l){2-3} 
                                  & 0.1 to 0.25       & Low sharpness         \\ \cmidrule(l){2-3} 
                                  & \textgreater 0.25 & Sharp                 \\ \bottomrule
\multirow{5}{*}{\textbf{Exposure quality}} & \textless -20     & Severely underexposed \\ \cmidrule(l){2-3} 
                                  & -20 to -10        & Underexposed          \\ \cmidrule(l){2-3} 
                                  & -10 to 10         & Well exposed          \\ \cmidrule(l){2-3} 
                                  & 10 to 20          & Overexposed           \\ \cmidrule(l){2-3} 
                                  & \textgreater 20   & Severely overexposed  \\ \bottomrule
\end{tabular}
\end{table}

\begin{table}[ht]
\caption{Monthly ad impressions, click-through rates and submission rates by gender and age}
\label{table:tableS4}
\begin{tabular}{@{}rccc@{}}
\toprule
\multicolumn{1}{l}{} &
  \textbf{\begin{tabular}[c]{@{}c@{}}Monthly Ad Views \\ Mean $\pm$ SD\end{tabular}} &
  \textbf{\begin{tabular}[c]{@{}c@{}}Ad click-through rate\\ Mean $\pm$ SD\end{tabular}} &
  \textbf{\begin{tabular}[c]{@{}c@{}}Submission Rate\\ Mean $\pm$ SD\end{tabular}} \\ \midrule
\multicolumn{1}{l}{} &
  \multicolumn{3}{c}{Gender} \\ \midrule
\multicolumn{1}{r|}{\textbf{Female}} &
  \multicolumn{1}{c|}{11,615$\pm$ 4968} &
  \multicolumn{1}{c|}{20.05\% $\pm$ 3.1} &
  \multicolumn{1}{c}{15.8\% $\pm$ 9.2} \\ \midrule
\multicolumn{1}{r|}{\textbf{Male}} &
  \multicolumn{1}{c|}{7,186 $\pm$ 2724} &
  \multicolumn{1}{c|}{13.8\% $\pm$ 4.1} &
  \multicolumn{1}{c}{14.1 $\pm$ 7.9} \\ \midrule
\multicolumn{1}{r|}{\textbf{Unknown}} &
  \multicolumn{1}{c|}{57,501 $\pm$ 25,313} &
  \multicolumn{1}{c|}{12.0 $\pm$ 11.9} &
  \multicolumn{1}{c}{8.9 $\pm$ 7.4} \\ \midrule
\multicolumn{1}{r|}{\textbf{All}} &
  \multicolumn{1}{c|}{77,739 $\pm$ 30,574} &
  \multicolumn{1}{c|}{13.6 $\pm$ 9.3} &
  \multicolumn{1}{c}{11.3 $\pm$ 8.5} \\ \midrule
\multicolumn{1}{r|}{\textbf{p value (two-sample t test)}} &
  \multicolumn{1}{c|}{0.044} &
  \multicolumn{1}{c|}{0.002} &
  \multicolumn{1}{c}{0.69} \\ \midrule
\multicolumn{1}{l}{} &
  \multicolumn{3}{c}{Age} \\ \midrule
\multicolumn{1}{r|}{\textbf{18–24}} &
  \multicolumn{1}{c|}{2,136.4 $\pm$ 895.7} &
  \multicolumn{1}{c|}{12.6 $\pm$ 3.8} &
  \multicolumn{1}{c}{21.3 $\pm$ 3.4} \\ \midrule
\multicolumn{1}{r|}{\textbf{25–34}} &
  \multicolumn{1}{c|}{3,834.9 $\pm$ 1,567.1} &
  \multicolumn{1}{c|}{13.8 $\pm$ 3.6} &
  \multicolumn{1}{c}{19.0 $\pm$ 8.6} \\ \midrule
\multicolumn{1}{r|}{\textbf{35–44}} &
  \multicolumn{1}{c|}{3,766.1 $\pm$ 1414} &
  \multicolumn{1}{c|}{15.0 $\pm$ 2.8} &
  \multicolumn{1}{c}{16.9 $\pm$ 7.8} \\ \midrule
\multicolumn{1}{r|}{\textbf{45–54}} &
  \multicolumn{1}{c|}{3,366.0 $\pm$ 1,495.8} &
  \multicolumn{1}{c|}{18.7 $\pm$ 2.1} &
  \multicolumn{1}{c}{15.0 $\pm$ 7.6} \\ \midrule
\multicolumn{1}{r|}{\textbf{54–55}} &
  \multicolumn{1}{c|}{3,161.0 $\pm$ 2,122.3} &
  \multicolumn{1}{c|}{21.1 $\pm$ 3.3} &
  \multicolumn{1}{c}{12.0 $\pm$ 7.5} \\ \midrule
\multicolumn{1}{r|}{\textbf{65+}} &
  \multicolumn{1}{c|}{2,371.6 $\pm$ 1944.5} &
  \multicolumn{1}{c|}{20.4 $\pm$ 4.9} &
  \multicolumn{1}{c}{9.4 $\pm$ 7.1} \\ \midrule
\multicolumn{1}{r|}{\textbf{Unknown}} &
  \multicolumn{1}{c|}{57,666. $\pm$ 23,699} &
  \multicolumn{1}{c|}{12.1 $\pm$ 11.1} &
  \multicolumn{1}{c}{9.0 $\pm$ 7.0} \\ \midrule
\multicolumn{1}{r|}{\textbf{All}} &
  \multicolumn{1}{c|}{77,739 $\pm$ 30,574} &
  \multicolumn{1}{c|}{13.6 $\pm$ 9.3} &
  \multicolumn{1}{c}{11.3 $\pm$ 8.5} \\ \midrule
\multicolumn{1}{r|}{\textbf{ANOVA p value}} &
  \multicolumn{1}{c|}{0.266} &
  \multicolumn{1}{c|}{$<$0.0001} &
  \multicolumn{1}{c}{0.037} \\ \midrule
\end{tabular}

\end{table}

\begin{table}[H]
\caption{Number and percentage of individuals reporting demographic and skin-condition-related information}
\label{table:tableS5}
\begin{tabular}{@{}r|c@{}}
\toprule
\textbf{Self-reported variable} & \textbf{Number of individuals, \% of dataset }                     \\ \midrule
\textbf{Age}                    & 2556, 50.76\%                      \\ \midrule
\textbf{Sex at birth}           & 2596, 51.56\%                      \\ \midrule
\textbf{Race/Ethnicity}         & 2648, 52.59\%                      \\ \midrule
\textbf{Fitzpatrick skin type}  & 2500, 49.65\%                      \\ \midrule
\textbf{Texture of condition}   & 4081, 81.05\%                      \\ \midrule
\textbf{Affected body part(s)}  & 4088, 81.19\%                      \\ \midrule
\textbf{Duration of condition}  & 4019, 79.82\%                       \\ \midrule
\textbf{Condition description}  & 3783, 75.13\%                      \\ \midrule
\textbf{Skin-related symptoms}  & 3775, 74.98\%                      \\ \midrule
\textbf{Other symptoms}         & 3508, 69.67\%                     \\ \bottomrule
\end{tabular}

\end{table}

\begin{table}[H]
\caption{SCIN dataset breakdown by age and sex}
\label{table:tableS6}
\begin{tabular}{@{}rccc@{}}
\toprule
\multicolumn{1}{c}{\textbf{}} & \multicolumn{1}{c|}{\textbf{SCIN dataset, self-reported (\%)}} & \multicolumn{1}{c|}{\textbf{US Population (\%)}} & \textbf{Residuals} \\ \midrule
\multicolumn{4}{c}{\textbf{Sex at birth*}}                                                                                               \\ \midrule
\multicolumn{1}{r|}{\textbf{Female}}                                   & \multicolumn{1}{c|}{66.72} & \multicolumn{1}{c|}{50.4} & 11.68  \\ \midrule
\multicolumn{1}{r|}{\textbf{Male}}                                     & \multicolumn{1}{c|}{33.28} & \multicolumn{1}{c|}{49.6} & -11.77 \\ \midrule
\multicolumn{1}{r|}{\textbf{Chi statistic, p value}} & \multicolumn{3}{c}{141.77, \textless{}0.001}                    \\ \midrule
\multicolumn{4}{c}{\textbf{Age group**}}                                                                                                 \\ \midrule
\multicolumn{1}{r|}{\textbf{18–28}}                                    & \multicolumn{1}{c|}{32.68}  & \multicolumn{1}{c|}{16.7} & 19.62  \\ \midrule
\multicolumn{1}{r|}{\textbf{30–39}}                                    & \multicolumn{1}{c|}{19.33} & \multicolumn{1}{c|}{13.5} & 9.58   \\ \midrule
\multicolumn{1}{r|}{\textbf{40–49}}                                    & \multicolumn{1}{c|}{20.62} & \multicolumn{1}{c|}{12.3} & 10.3   \\ \midrule
\multicolumn{1}{r|}{\textbf{50–59}}                                    & \multicolumn{1}{c|}{15.03} & \multicolumn{1}{c|}{12.7} & 3.35   \\ \midrule
\multicolumn{1}{r|}{\textbf{60–69}}                                    & \multicolumn{1}{c|}{8.98}  & \multicolumn{1}{c|}{12.0} & -4.26  \\ \midrule
\multicolumn{1}{r|}{\textbf{70–79}}                                    & \multicolumn{1}{c|}{2.81}  & \multicolumn{1}{c|}{7.7}  & -8.89  \\ \midrule
\multicolumn{1}{r|}{\textbf{80+}}                                      & \multicolumn{1}{c|}{0.55}  & \multicolumn{1}{c|}{3.7}  & -8.28  \\ \midrule
\multicolumn{1}{r|}{\textbf{Chi statistic, p value}} & \multicolumn{3}{c}{279.87, \textless{}0.001}                    \\ \bottomrule
\end{tabular}
\small *US population sex composition from \href{https://www.census.gov/quickfacts/fact/table/US/PST045222}{United States Census Bureau Quickfacts 2022}, race/ethnicity composition can be obtained from the same source. Race and ethnicity categories are not comparable enough for statistical analysis. **Age group composition from \href {https://www2.census.gov/programs-surveys/demo/tables/age-and-sex/2022/age-sex-composition/2022agesex_table1.xlsx}{US census bureau table}, 2022
\end{table}

\begin{table}[H]
\caption{SCIN dataset condition duration}
\label{table:tableS7}
\begin{tabular}{@{}r|c@{}}
\toprule
\textbf{Condition Duration} & \textbf{\% of SCIN dataset$^*$} \\ \cmidrule(lr){1-2}
\textbf{$<$ 1 day} & 19.40 \\ \cmidrule(lr){1-2}
\textbf{1 – 7 days} & 34.56 \\ \cmidrule(lr){1-2}
\textbf{7 – 28 days} & 23.86 \\ \cmidrule(lr){1-2}
\textbf{1 – 3 months} & 7.86 \\ \cmidrule(lr){1-2}
\textbf{3 months – 1 year} & 5.94 \\ \cmidrule(lr){1-2}
\textbf{1 – 5 years} & 5.05 \\ \cmidrule(lr){1-2}
\textbf{5 – 10 years} & 2.33 \\ \cmidrule(lr){1-2}
\textbf{$>$ 10 years} & 0.97 \\ \bottomrule
\end{tabular}

\small $^*$Includes contributions where condition duration was reported (N=4031)
\end{table}

\begin{table}[H]
\caption{SCIN dataset dermatology condition distribution}
\label{table:tableS8}
\begin{tabular}{@{}l|l|c@{}}
\toprule
\textbf{Dermatology Condition Category} & \textbf{Dermatology Condition Subcategory} & \textbf{\% of SCIN dataset*} \\ \cmidrule(lr){1-1} \cmidrule(lr){2-2} \cmidrule(lr){3-3}
Eruptions & Inflammatory Eruption & 35.55 \\ \cmidrule(lr){2-3} & Other Eruption & 14.41 \\ \cmidrule(lr){2-3} & Reactive Eruption & 3.13 \\ \cmidrule(lr){2-3} & Drug Eruption & 2.06 \\ \midrule
Cutaneous Infection & Cutaneous Infection & 21.81 \\ \midrule
Contact Dermatitis & Contact Dermatitis & 10.05 \\ \midrule
Vascular condition & Vascular condition & 2.24 \\ \midrule
Pigmentary Disorder & Pigmentary Disorder & 0.75 \\ \midrule
Nail Condition & Nail Condition & 0.25 \\ \midrule
Hair Disorder & Hair Disorder & 0.04 \\ \midrule
Benign Neoplasms & Benign Neoplasm & 2.53 \\ \cmidrule(lr){2-3} & Benign Neoplasm - Reactive & 1.32 \\ \midrule
Malignant and pre-malignant neoplasms & Malignant Neoplasm & 0.85 \\ \cmidrule(lr){2-3} & Possibly Premalignant Neoplasm & 0.5 \\ \midrule
Others & Others & 2.35 \\ \bottomrule
\end{tabular}

\small $^{*}$Includes contributions that could be labeled as one of 419 SNOMED condition categories (N=2835, See Methods)
\end{table}

\begin{table}[H]
\caption{Dermatology condition distribution: Existing dermatology datasets and SCIN}
\label{table:tableS9}
\begin{tabulary}{\linewidth}{@{}LCCCCCCCCCCC@{}}
      \toprule
   & &
  Eruptions &
  Cutaneous infection &
  Contact Dermatitis &
  Vascular &
  Pigment Disorder &
  Ulcers and Blisters &
  Benign neoplasm &
  Malignant or premalignant &
  Nail or Hair &
  Other \\
      \midrule
      {\textbf{SCIN}} &  &
  \multicolumn{1}{c}{56.43} &
  \multicolumn{1}{c}{21.85} &
  \multicolumn{1}{c}{10.5} &
  \multicolumn{1}{c}{2.24} &
  \multicolumn{1}{c}{0.75} &
  \multicolumn{1}{c}{0.32} &
  \multicolumn{1}{c}{3.85} &
  \multicolumn{1}{c}{1.35} &
  \multicolumn{1}{c}{0.29} &
  \multicolumn{1}{c}{2.42}  \\\addlinespace
      {\textbf{Fitzpatrick17k}} & &
  \multicolumn{1}{c}{54.63} &
  \multicolumn{1}{c}{6.21} &
  \multicolumn{1}{c}{3.12} &
  \multicolumn{1}{c}{1.7} &
  \multicolumn{1}{c}{1.53} &
  \multicolumn{1}{c}{1.51} &
  \multicolumn{1}{c}{13.04} &
  \multicolumn{1}{c}{13.89} &
  \multicolumn{1}{c}{0.0} &
  \multicolumn{1}{c}{4.38} \\\addlinespace
        {\textbf{DDI}} & &
  \multicolumn{1}{c}{2.28} &
  \multicolumn{1}{c}{1.66} &
  \multicolumn{1}{c}{0.0} &
  \multicolumn{1}{c}{0.76} &
  \multicolumn{1}{c}{0.46} &
  \multicolumn{1}{c}{0.0} &
  \multicolumn{1}{c}{66.01} &
  \multicolumn{1}{c}{26.68} &
  \multicolumn{1}{c}{0.3} &
  \multicolumn{1}{c}{1.83} \\\addlinespace
  {\textbf{ISIC 2020}} & &
  \multicolumn{1}{c}{0.0} &
  \multicolumn{1}{c}{0.0} &
  \multicolumn{1}{c}{0.0} &
  \multicolumn{1}{c}{0.0} &
  \multicolumn{1}{c}{0.0} &
  \multicolumn{1}{c}{0.0} &
  \multicolumn{1}{c}{98.24} &
  \multicolumn{1}{c}{1.76} &
  \multicolumn{1}{c}{0.0} &
  \multicolumn{1}{c}{0.0}\\\addlinespace
  {\textbf{ISIC 2017}} & &
  \multicolumn{1}{c}{0.0} &
  \multicolumn{1}{c}{0.0} &
  \multicolumn{1}{c}{0.0} &
  \multicolumn{1}{c}{0.0} &
  \multicolumn{1}{c}{0.0} &
  \multicolumn{1}{c}{0.0} &
  \multicolumn{1}{c}{92.19} &
  \multicolumn{1}{c}{7.81} &
  \multicolumn{1}{c}{0.0} &
  \multicolumn{1}{c}{0.0}\\\addlinespace
  {\textbf{PH2}} & &
  \multicolumn{1}{c}{0.0} &
  \multicolumn{1}{c}{0.0} &
  \multicolumn{1}{c}{0.0} &
  \multicolumn{1}{c}{0.0} &
  \multicolumn{1}{c}{0.0} &
  \multicolumn{1}{c}{0.0} &
  \multicolumn{1}{c}{80.0} &
  \multicolumn{1}{c}{20.0} &
  \multicolumn{1}{c}{0.0} &
  \multicolumn{1}{c}{0.0}\\\addlinespace
  {\textbf{HAM10k}} & &
  \multicolumn{1}{c}{0.0} &
  \multicolumn{1}{c}{0.0} &
  \multicolumn{1}{c}{0.0} &
  \multicolumn{1}{c}{0.0} &
  \multicolumn{1}{c}{0.0} &
  \multicolumn{1}{c}{0.0} &
  \multicolumn{1}{c}{80.21} &
  \multicolumn{1}{c}{19.79} &
  \multicolumn{1}{c}{0.0} &
  \multicolumn{1}{c}{0.0}\\\addlinespace
  {\textbf{SKINL2}} & &
  \multicolumn{1}{c}{0.0} &
  \multicolumn{1}{c}{0.0} &
  \multicolumn{1}{c}{0.0} &
  \multicolumn{1}{c}{1.76} &
  \multicolumn{1}{c}{0.0} &
  \multicolumn{1}{c}{0.0} &
  \multicolumn{1}{c}{74.12} &
  \multicolumn{1}{c}{20.59} &
  \multicolumn{1}{c}{0.0} &
  \multicolumn{1}{c}{3.53}\\\addlinespace
  \textbf{PAD-UFES} & &
  0.0 &
  0.0 &
  0.0 &
  0.0 &
  0.0 &
  0.0 &
  20.84 &
  79.16 &
  0.0 &
  0.0\\\bottomrule
\end{tabulary}
\end{table}

\begin{table}[H]
\caption{Fitzpatrick skin type distribution: Existing dermatology datasets and SCIN}
\label{table:tableS10}
\resizebox{\textwidth}{!}{%
\begin{tabular}{@{}rcccccc@{}}
\toprule
\multicolumn{1}{l}{}      & \textbf{Type 1 (\%)} & \textbf{Type 2 (\%)} & \textbf{Type 3 (\%)} & \textbf{Type 4 (\%)} & \textbf{Type 5 (\%)} & \textbf{Type 6 (\%)} \\ \midrule
\textbf{SCIN sFST}             & 8.64                 & 24.92                & 30.39                & 19.63                & 9.84                & 6.57                 \\
\textbf{SCIN eFST}             & 7.55                 & 40.21                & 32.76                & 13.64               & 5.27                & 0.57                 \\
Fitzpatrick17k & 23.17                & 33.98                & 18.39                & 14.24                & 7.0                  & 3.23                 \\
PH2            & 50.0                 & 50.0                 & 0.0                  & 0.0                  & 0.0                  & 0.0                  \\
SKINL2         & 0.54                 & 56.33                & 42.86                & 0.27                 & 0.0                  & 0.0                  \\
PAD-UFES       & 10.24                & 58.63                & 26.24                & 4.15                 & 0.67                 & 0.07                 \\ \bottomrule
\end{tabular}%
}
\end{table}

\begin{table}[H]
\caption{Estimated Monk Skin Tone distribution for the SCIN dataset}
\label{table:tableS11}
\begin{tabular}{@{}r|c@{}|c@{}}
\toprule
\textbf{eMST} & \textbf{US labelers(\%)}    & \textbf{India labelers (\%)}     \\ \midrule
\textbf{1}                    & 4.11       & 10.44         \\ \midrule
\textbf{2}           & 43.19     & 36.52         \\ \midrule
\textbf{3}         & 31.83     & 25.78         \\ \midrule
\textbf{4}  & 12.4     & 12.66         \\ \midrule
\textbf{4}   & 4.68      & 6.63           \\ \midrule
\textbf{6}  & 2.03      & 4.33          \\ \midrule
\textbf{7}  & 1.12      & 2.46          \\ \midrule
\textbf{8}  & 0.54      & 0.96          \\ \midrule
\textbf{9}  & 0.1       & 0.19          \\ \midrule
\textbf{10}         & 0.0       & 0.04          \\ \bottomrule
\end{tabular}
\end{table}

\subsection*{Supplementary Figures}
\setcounter{figure}{0}
\renewcommand{\thefigure}{S\arabic{figure}}

\begin{figure}[H]
\caption{Mock-up of the SCIN web application showing the submission flow}
\label{fig:figS1}
\includegraphics[width=\textwidth]{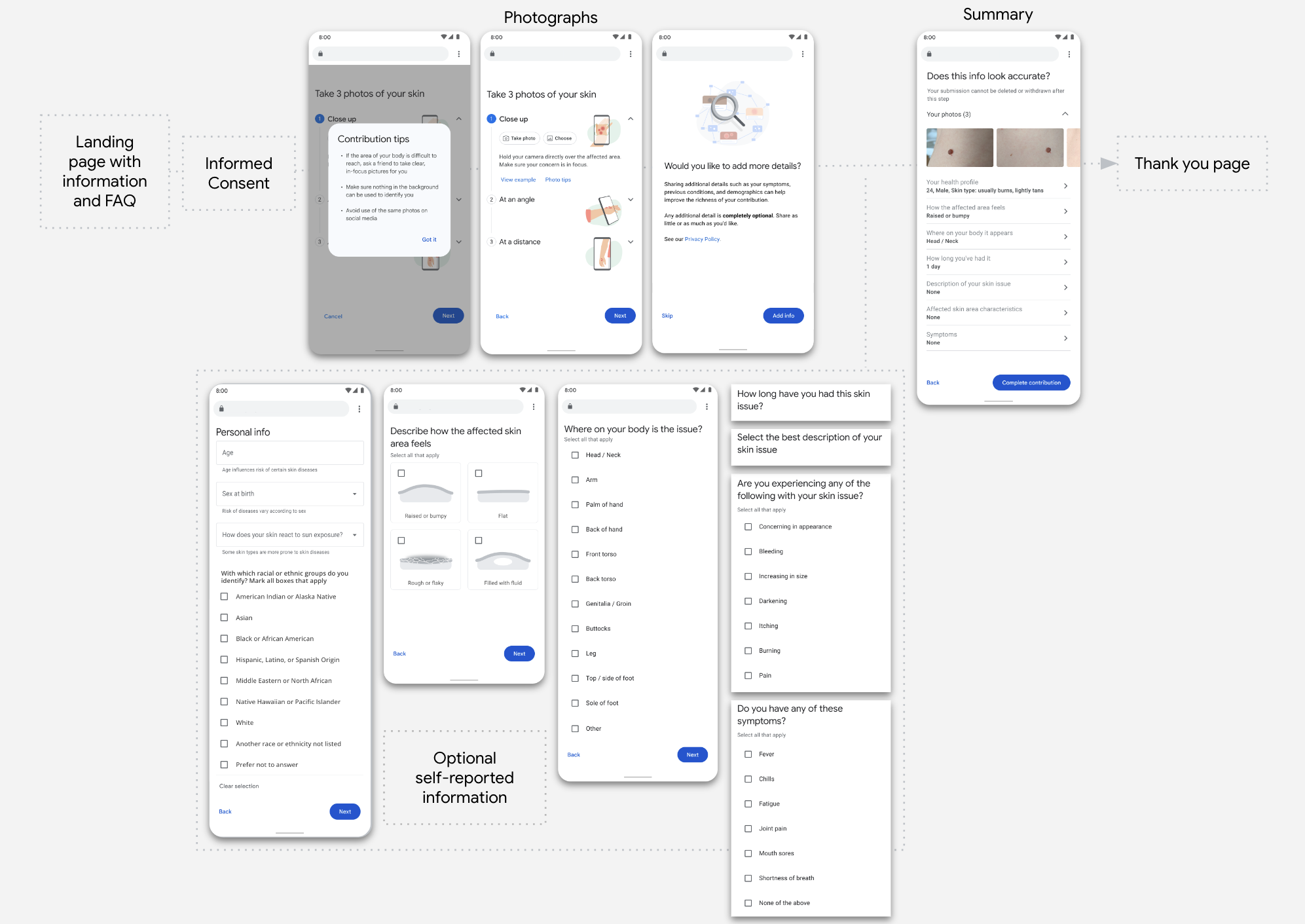}
\end{figure}

\begin{figure}[H]
\caption{Comparisons between A,B: Dermatologist-labeled estimated Fitzpatrick Skin Type (eFST) and estimated Monk Skin tone (eMST) label, C,D: Self-reported FST (sFST) and eMST label E: eFST and sFST}
\label{fig:figS2}
\includegraphics[width=\textwidth]{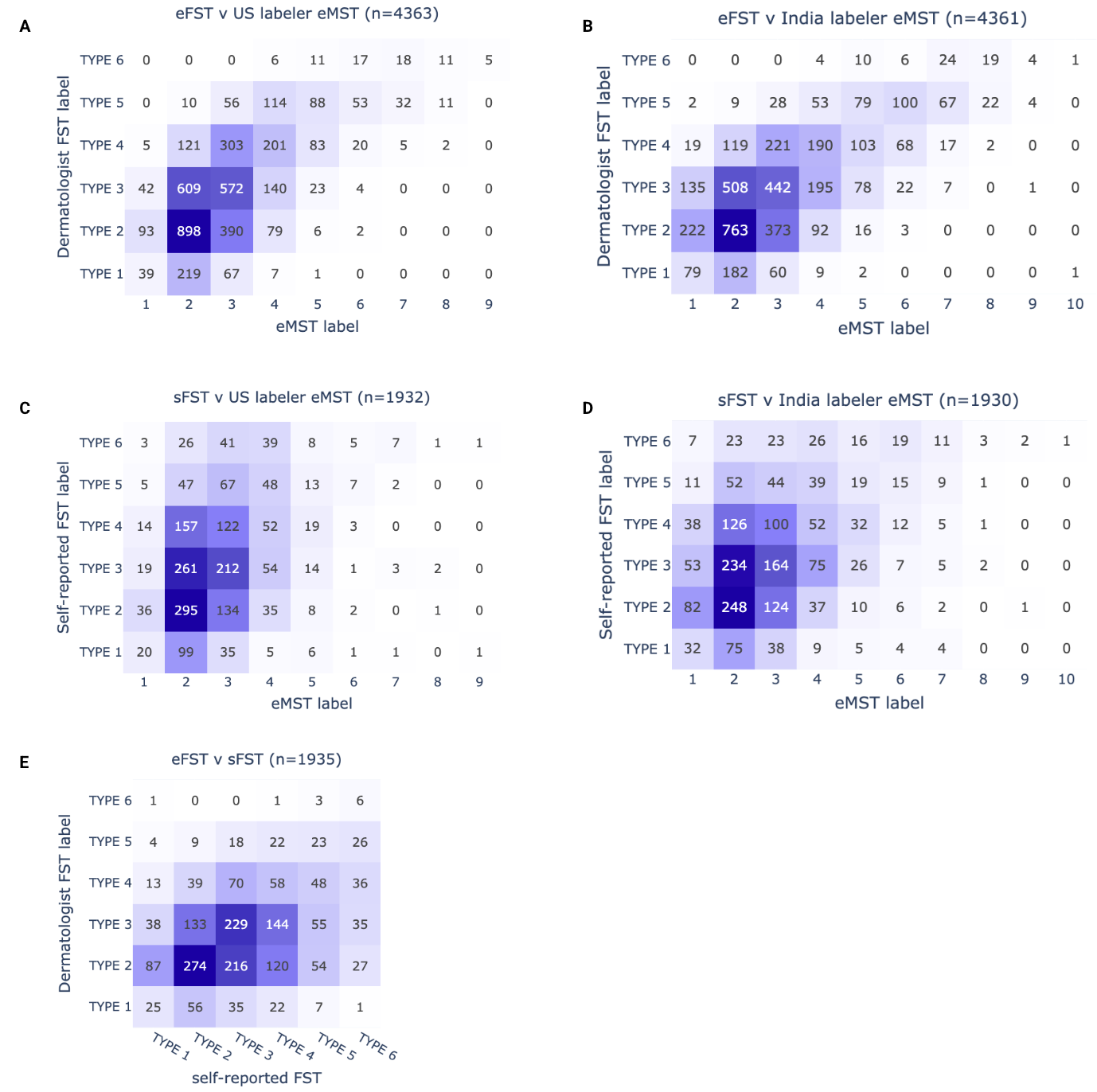}
\end{figure}

\end{document}